\documentclass[aps,pra,nofootinbib,floats,floatfix,twocolumn,superscriptaddress]{revtex4-2}
\usepackage{latexsym}
\usepackage{amsmath,amssymb,amsthm,thmtools}
\usepackage{mathtools,mathrsfs}
\usepackage{amsbsy}
\usepackage{soul}
\usepackage[caption=false]{subfig}
\usepackage{bbold}

\usepackage[utf8]{inputenc}

\usepackage[pdftex]{graphicx}
\graphicspath{{./figures/}}

\usepackage[usenames,dvipsnames,table]{xcolor}
\usepackage{float}

\usepackage[colorlinks=true]{hyperref}
\usepackage[nameinlink]{cleveref}
\hypersetup{
  colorlinks   = true, %Colours links instead of ugly boxes
  urlcolor     = green!80!black, %Colour for external hyperlinks
  linkcolor    = blue, %Colour of internal links 	q
  citecolor    = red!80!black %Colour of citations
}

\usepackage{graphicx}

\usepackage[normalem]{ulem}

\usepackage{microtype}

\usepackage{physics}
\usepackage{parskip}
\usepackage{easyReview}

\newcommand{\calE}{\mathcal{E}}
\newcommand{\calF}{\mathcal{F}}

\newcommand{\calS}{\mathcal{S}}

\newcommand{\parTitle}[1]{\noindent\emph{#1} ---}
\begin{document}

\title{
The importance of using the averaged mutual information when quantifying quantum objectivity
}

\author{Diana A.~Chisholm}
\affiliation{Universit\`a degli Studi di Palermo, Dipartimento di Fisica e Chimica -- Emilio Segr\`e,\\ via Archirafi 36, I-90123 Palermo, Italy}
\affiliation{Centre for Theoretical Atomic, Molecular, and Optical Physics, School of Mathematics and Physics, Queen’s University Belfast, BT7 1NN, United Kingdom}
\author{Luca Innocenti}
\affiliation{Universit\`a degli Studi di Palermo, Dipartimento di Fisica e Chimica -- Emilio Segr\`e,\\ via Archirafi 36, I-90123 Palermo, Italy}
\author{G.~Massimo Palma}
\affiliation{Universit\`a degli Studi di Palermo, Dipartimento di Fisica e Chimica -- Emilio Segr\`e,\\ via Archirafi 36, I-90123 Palermo, Italy}
\affiliation{NEST, Istituto Nanoscienze-CNR, Piazza S.~Silvestro 12, 56127 Pisa, Italy}

\begin{abstract}
    In the context of quantum objectivity, a standard way to quantify the classicality of a state is via the mutual information between a system and different fractions of its environment.
    Many of the tools developed in the relevant literature to quantify quantum objectivity via quantum mutual information rely on the assumption that information about the system leaks symmetrically into its environment.
    In this work, we highlight the importance of taking this assumption into account, and in particular, analyse how taking non-averaged quantum mutual information as a quantifier of quantum objectivity can be severely misleading whenever information about the system is encoded into the environment in a non-homogeneous way. On the other hand, the averaged mutual information always provides results with a clear operative interpretation.
\end{abstract}

\maketitle

\section{Introduction}

The study of quantum objectivity aims at bettering our understanding of the quantum-to-classical transition~\cite{ollivier_objective_2004,ollivier_environment_2005}.
In this context, a recognized hallmark of classical systems is the ability for multiple observers to reach a consensus on the state of a physical system, and thus access information without disrupting it.
A standard approach to quantifying quantum objectivity involves analyzing the behaviour of the quantum mutual information (QMI) between system and different environmental fractions~\cite{blume-kohout_quantum_2006, zurek_quantum_2009,le_strong_2019}. We will refer to this approach as \emph{QD-objectivity}.
Another possibility is to witness objectivity by quantifying how close the system is to having a so-called spectrum broadcast structure (SBS)~\cite{korbicz_objectivity_2014,horodecki_quantum_2015}.
These two approaches were recently showed to reflect two related, but distinct, features of quantum objectivity. Namely, QD-objectivity is related to the probability that a number of observers having access to different pieces of the environment will agree on the state of the system, whereas SBSs are related to the number of times the system is encoded into its environment, regardless of the hardness of actually observing such correlations~\cite{chisholm2023meaning}.

Here, we reconsider the way objectivity is discussed, in the framework of quantum Darwinism (QD), via QMIs, leveraging the recently introduced operative definitions of redundancy and consensus~\cite{chisholm2023meaning} to achieve a deeper insight into what precisely the analyzed quantities correspond to.
We will show, in particular, that in scenarios involving asymmetric environments, which occurs frequently in practice~\cite{milazzo_role_2019, pleasance2017application, riedel2010quantum, chisholm_witnessing_2021, garcia-perez_decoherence_2020, le_objectivity_2018, ciampini_experimental_2018}, using the non-averaged QMI can lead to severe misinterpretations of the objectivity features of a system. The averaged QMI on the other hand, which is tightly connected to the notion of consensus, always provides consistent results. We do this by analysing an ideal, easy to interpret, scenario as well as a more realistic one.
We also analyse the properties of the non-averaged QMI and conclude that, while it is a misleading quantifier of quantum obectivity, it still provides interesting informations regarding the stucture of the correlations between system and environment.

\parTitle{Outline}
The rest of the paper is structured as follows: in \cref{sec:quantifying_objectivity} we outline the framework of QD to measure quantum objectivity and we introduce the averaged QMI, we then present the concepts of \emph{redundancy} and \emph{consensus}, which will be useful in our analysis.
In \cref{sec:asymm_environments} we provide a detailed analysis of the behaviour of averaged and non-averaged QMI as a function of the considered environmental fraction, in the case of asymmetric information encoding into the environment. We will make use of a specific but illustrative example where some of the environment is uncorrelated with the system.
In \cref{sec:random_correlations} we will study a more generic case of information encoding, with each environmental constituent having a random degree of correlation with the system. We will show that the results provided in the paper apply to this case as well.
In \cref{sec:non-averaged-qmi} we will discuss some general properties of the non-averaged QMI, and how they can help, in the context of quantum objectivity, in inferring relevant information about the overall structure of the system-environment correlations.
In the last Section, we will make our final remarks.

\section{Quantifying quantum objectivity}
\label{sec:quantifying_objectivity}
In this section we briefly overview the approach to quantify quantum objectivity, via the framework of QD~\cite{blume-kohout_quantum_2006, zurek_quantum_2009, le_strong_2019}, and the operative definitions of redundancy and consensus~\cite{chisholm2023meaning}, which will prove uesful in the later sections.

\parTitle{QD-objectivity}
Consider a system-environment state $\rho_{\calS\calE}$ shared between a system $\calS$ and an environment $\calE=\bigotimes_{i=1}^n\calE_i$ composed of $n$ elementary constituents $\calE_i$.
Although the formalism is straightforwardly generalizable to arbitrary dimensions, we focus here on the most common scenario of many-qubit systems, where $\calS\simeq\calE_k\simeq \mathbb{C}^2$ for all $k$.
Quantum Darwinism (QD) quantifies its ``objectivity'' studying how the QMI between $\calS$ and a subset of $l$ environmental qubits changes with $l$.
More formally, we thus consider the QMI $I(\calS:\calE_K)$, with $\calE_K\equiv\bigotimes_{k\in K}\calE_k$ an $l$-qubit subsystem, $K\subseteq\{1,...,n\}$, and $|K|=l\le n$.
For practical reasons, rather than considering the relation $|K|\mapsto I(\calS:\calE_K)$, it is standard practice to reparametrize $l$ via the environmental fraction $f\in[0,1]$ it corresponds to.
We will then denote with
\begin{equation}
    I(\calS:\calE_f) \equiv
    S(\rho_\calS) + S(\rho_{\calE_f})
    - S(\rho_{\calS\calE_f})
\end{equation}
the QMI between $\calS$ and a subset $\calE_f$ of $\dim(\calE_f)=f \dim(\calE)$ environmental qubits~\footnote{In this context, it is generally assumed that $f \dim(\calE)\in\mathbb{N}$. For lager fractions and system sizes, this subtlety becomes unimportant.}.
The state $\rho_{\calS\calE}$ is then said to be \emph{QD-objective} if $I(\calS:\calE_f)\simeq S(\rho_\calS)$ for all $f\in(f_0, 1-f_0)$ for a suitable threshold $f_0\in(0,1)$.
% and for all subspaces $\calE_f\le\calE$ with $\dim(\calE_f)=f\dim(\calE)$.
%
If this condition is met, any observer having access to a fraction $\calE_{f_0}$ will be able to recover information about $\calS$, and therefore any partitioning of the environment into $\dim(\calE)/\dim(\calE_{f_0})=1/f_0$ disjoint fractions results in $1/f_0$ observers which can agree on the state of the system.

An underying assumption in the above definition of QD-objectivity is that the QMI $I(\calS:\calE_f)$ is well-defined by only specifying the \emph{size} of the environmental fraction $\calE_f$.
This assumption holds whenever the dynamic is driven by an interaction Hamiltonian of the form $H_I=A^\calS\otimes\sum_k B^{\calE_k}$, and thus the system correlates identically with each environmental fraction of a given size~\cite{mironowicz_monitoring_2017,lorenzo_anti-zeno-based_2020,megier_correlations_2022, touil_eavesdropping_2022, roszak_entanglement_2019, roszak_glimpse_2020, cakmak_quantum_2021, schwarzhans2023quantum}
Many situations of interest, however, do not sport this kind of symmetry~\cite{milazzo_role_2019, pleasance2017application, riedel2010quantum, chisholm_witnessing_2021, garcia-perez_decoherence_2020, le_objectivity_2018, ciampini_experimental_2018}, and thus the dependence of the QMI $I(\calS:\calE_f)$ on $f$ might depend on which specific sequence of fractions $f\mapsto \calE_f$ is considered, as will be discussed in detail in the later sections.

\parTitle{Averaged QMI}
In such asymmetric scenarios, $I(\calS:\calE_f)$ is not a well-defined function of $f\in[0,1]$, and the definitions given above for QD-objectivity may produce results that depend on the specific choice of increasing sequence of environmental fractions $f\mapsto \calE_f$.
One possible way to overcome this issue is to define QD-objectivity in terms of the \emph{averaged} QMI instead~\cite{blume-kohout_simple_2005, Zwolak2017Redundancy}.
%We thus define an averaged QMI $\tilde I_l(\calS:\calE)$ as the average of $I(\calS:\calF)$ among all possible $l$-qubit subsystems $\calF$.
We define the averaged QMI as
\begin{equation}\label{eq:averaged_QMI}
    \tilde I(\calS:\calE_f) = \langle I(\calS:\calE_f) \rangle_{\calE_f} \equiv
    \frac{1}{\binom{d}{fd}} \sum_{\calE_f\le \calE} I(\calS:\calE_f),
\end{equation}

where the sum is extended over all subspaces $\calE_f$ of dimension $fd$, and $d\equiv\dim(\calE)$.

\parTitle{Definitions of redundancy and consensus} 
The notions of ``redundancy'' and ``consensus'' of a state $\rho_{\calS\calE}$ were recently defined as follows~\cite{chisholm2023meaning}:
\begin{itemize}
    \item Redundancy is the size of the finest possible environment partition such that each individual fraction is maximally correlated with the system. In other words, the redundancy of $\rho_{\calS\calE}$ is $n$ if there is a nontrivial partition $\calE=\bigotimes_{i=1}^n \calE_i$ such that $I(\calS:\calE_i)\simeq S(\rho_{\calS})$ for all $i$, and there is no finer such partition~\footnote{In the original definition, the accessible information is used instead of the quantum mutual information, to avoid problems related to the presence of discord. In this work however we will mostly remain agnostic to this point, and the definitions remain consistent regardless of the type of mutual information used.}.
    \item Consensus, on the other hand, quantifies the likelihood of actually observing the information redundantly encoded into the environment. 
    It is the largest number of fractions such that, when chosen randomly, each of them will be fully correlated with the system, with probability close to one.
    The consensus is thus always upper bounded by the redundancy, with the two notions coinciding precisely in the cases of completely symmetric information encoding of $\calS$ into its environment. Consensus can be measured as $1/f_0$, where $f_0$ satisfies the condition that $\tilde I(\calS:\calE_{f_0})\simeq S(\rho_\calS)$.
\end{itemize}

\section{Asymmetric environments}
\label{sec:asymm_environments}

\begin{figure}[!]
    \centering
    \includegraphics[width=\columnwidth]{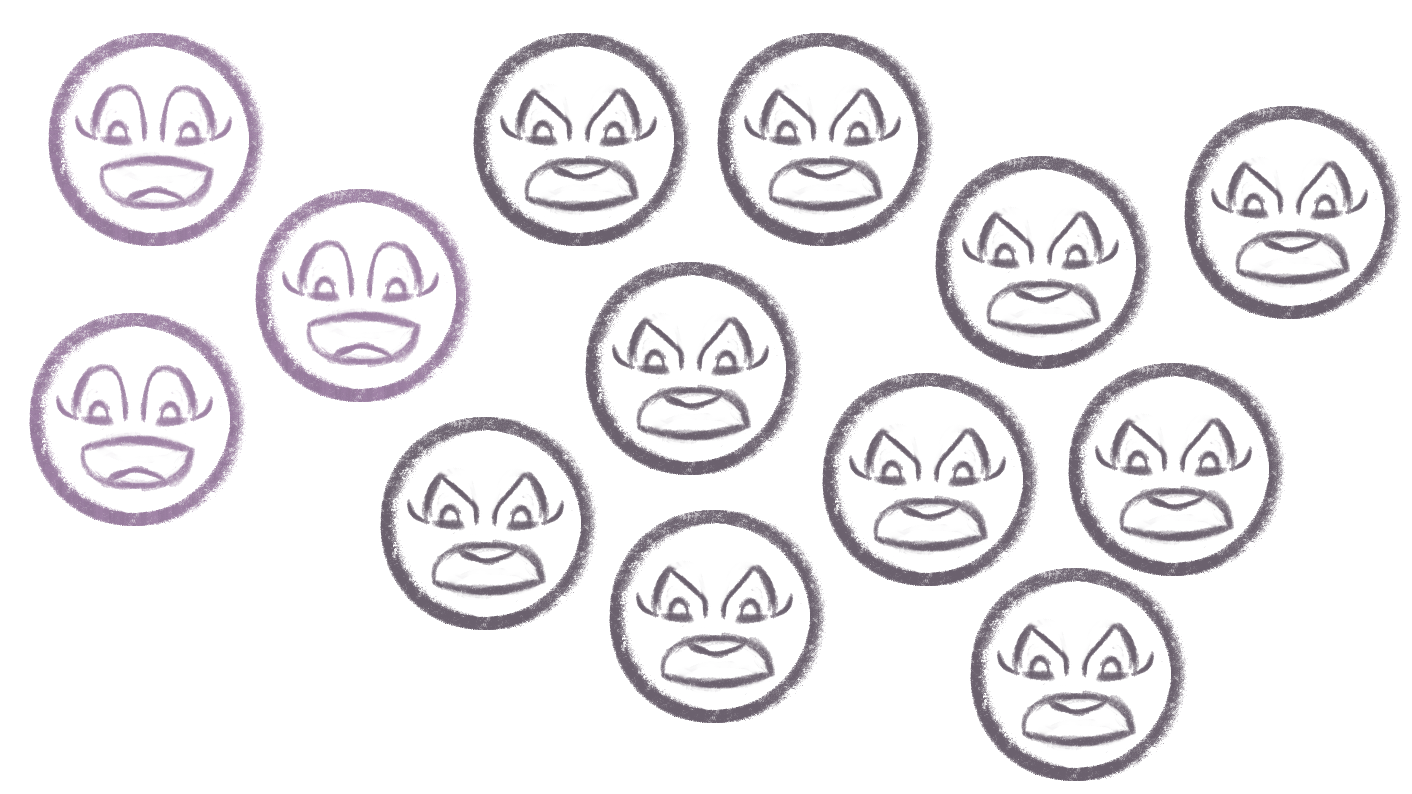}
    \caption{\textbf{Pictorial representation of the scenario described by \cref{eq:first_state}}: Only a small part of the environmental qubits are actually correlated with the system (happy faces), while the rest of the environment is entirely uncorrelated with it (angry faces). When observers perform measurements of their respective environmental fractions, some of the observers may end up only having access to uncorrelated qubits, and therefore be unable to agree with the other observers.}
    \label{fig:drawing}
\end{figure}

In this section we present a class of examples highlighting the departure between the notions of redundancy and consensus~\cite{chisholm2023meaning} even in situations with entirely local information encoding of the system into its environment.

%\parTitle{Redundancy of garbage states}
Consider $(N+1)$-qubit states of the form
\begin{equation}
    \label{eq:first_state}
    \ket{\Psi}=\frac{1}{\sqrt{2}}(\ket{0}_{\mathcal{S}}\otimes\ket{0}^{\otimes m}+\ket{1}_{\mathcal{S}}\otimes\ket{1}^{\otimes m})\otimes\ket{\phi}^{\otimes (N-m)},
\end{equation}
where the system is a single qubit, and the environment is made of $N$ qubits, $m<N$ of which are perfectly correlated with the system, while the remaining $N-m$ are entirely uncorrelated to the system (``junk qubits'').
%\cLI{instead of using $\ket+$ as junk qubits, I'd write it as $\ket{\phi}^{\otimes (N-m)}$ and say that the states of these is irrelevant}
Measuring any one of the $m$ correlated qubits is sufficient to recover the state of $\calS$, and thus information about the system is encoded redundantly into $m$ different environmental qubits. Following the definition of redundancy discussed in~\cref{sec:quantifying_objectivity}, the redundancy of $\ket\Psi$ is thus precisely $m$.
\begin{figure}[!]
    \centering
    \includegraphics[width=\columnwidth]{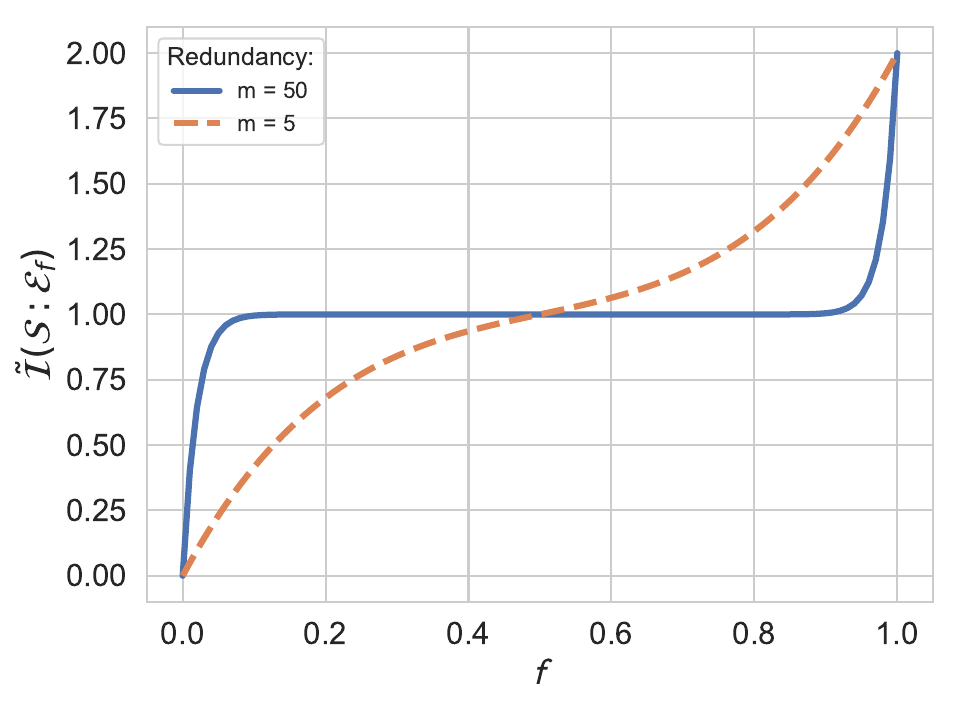}
    \caption{\textbf{Averaged mutual information as a function of the fraction size}. The environment consists of some qubits perfectly correlated with the system, and others entirely uncorrelated (``junk'' qubits). We compare the case in which there are $m=50$ correlated qubits, where we can see a clear signature of objectivity (consensus $11$), with the case in which there are only $m=5$ correlated qubits, where objectivity does not emerge (consensus $1$). In both cases, for every correlated qubit there are 20 uncorrelated ones. The value of redundancy is precisely the number of correlated qubits.
    }
    \label{fig:ave_good}
\end{figure}
On the other hand, since only some environmental constituents are correlated with the system, there is a high likelihood that a random partitioning of the environment leads to some observers only holding uncorrelated qubits. This results in consensus being strictly less than redundancy.

\parTitle{Calculation of averaged QMI}
In order to compute the averaged QMI, as  defined in~\cref{eq:averaged_QMI}, we consider the QMI $I(\calS:\calF)$ computed over different subsets of $l$ qubits, spanning a subspace $\calF\le\calE$ of the environment:
\begin{enumerate}
    \item If $\calF$ contains no correlated qubit, then $I(\calS:\calF)=0$. For this to be possible we must have $l\le m$, in which case there are $\binom{N-m}{l}$ possible ways to pick $l$ such qubits.
    \item If $\calF$ contains all the correlated qubits, potentially along with additional uncorrelated ones, then $I(\calS:\calF)=2S(\rho_\calS)$.
    This requires $l\ge m$, and there are $\binom{N-m}{l-m}$ possible ways to pick $l$ such qubits, corresponding to the number ways to pick $l-m$ qubits among the $N-m$ uncorrelated ones.
    \item If $\calF$ contains at least one, but not all, correlated qubits, then $I(\calS:\calF)=S(\rho_\calS)$.
    %These are the events where one picks at least one correlated and at least one uncorrelated qubit.
    This requires $1<l < N$, and there are
    $\binom{N}{l}-\binom{N-m}{l}-\binom{N-m}{l-m}$ such events. This figure is found observing that there are $\binom{N-m}{l}$ ways to choose \textit{only} uncorrelated qubits, and $\binom{N-m}{l-m}$ ways to choose \textit{all} of the correlated qubits, and we are considering the complement of these two disjoint events.
\end{enumerate}
The averaged QMI is therefore
\begin{equation}
    \tilde{I}(\calS:\calE_f)=\left(1-\frac{\binom{N-m}{fN}+\binom{N-m}{fN-m}}{\binom{N}{fN}}\right)S(\rho_\calS),
    \label{eq:combinatorial_QMI}
\end{equation}
where we reparametrized the fractions as $f\equiv l/N$.

The behaviour of $\tilde{I}(\calS:\calE_f)$ as a function of $f$ is shown in~\cref{fig:ave_good} for the cases with $N=1000$ and $m=50$, and $N=100$ and $m=5$, both of which corresponding to having only one in twenty qubits correlated with the system.
As evident from the figure, higher values of redundancy $m$ correspond to curves in which the typical plateau of objective states~\cite{zurek_quantum_2009} at $\tilde I(\calS:\calE_f)\simeq S(\rho_\calS)$ is reached for smaller fractions $f$.
On the other hand, when the redundancy is small --- for example $m=5$ in the figure --- there is no plateau, meaning that the state is not objective according to QD. In the formalism of~\cite{chisholm2023meaning}, this corresponds to a lack of consensus between observers among which the environment was randomly distributed.
More precisely, by choosing a threshold value of $0.99S(\rho_\calS)$, we obtain a consensus value of $11$ when the redundancy is $50$, and $1$ when the redundancy is $5$. A consensus value of 1 implies that only one observer is able to infer information about the system, and therefore no objectivity is achieved.

The QMI plots resulting from \cref{eq:combinatorial_QMI} remain largely unchanged when changing $N$, and only change significantly when varying $m$.
We thus see that objectivity emerges as a direct consequence of the number of correlated environment qubits, regardless of the amount of ``junk'' qubits.

\begin{figure}[tbh]
    \centering
    \includegraphics[width=\columnwidth]{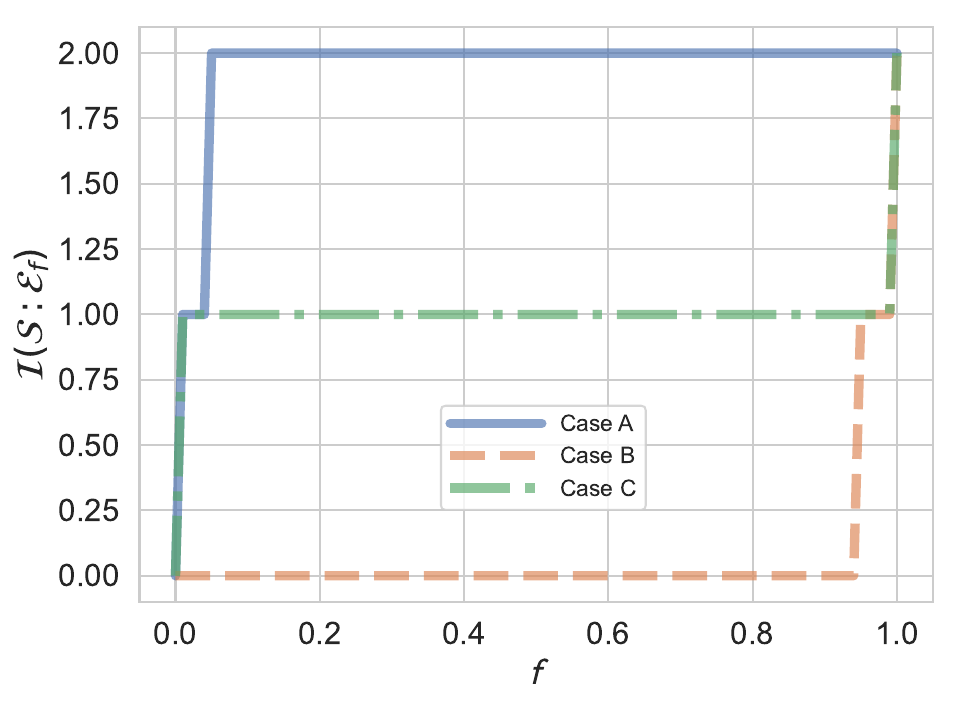}
    \caption{\textbf{Mutual information as a function of the fraction size}. 
    All plots refer to the same system-environment state, where however no average was performed. 
    In case A every fraction of size $fN$ contains only correlated qubits, in case B every fraction of size $fN$ contains only uncorrelated qubits, and in case C every fraction contains at least one correlated qubit, but at least one of the correlated qubits is always left out of each fraction.
    Not performing the average results in outcomes that are contradicting among themselves and misleading compared to the physical scenario.
    Since the ratio between correlated and uncorrelated qubits is the same, the presented plots are the same whether the total number of environmental qubits is $N=100$ or $N=1000$.
    }
    \label{fig:ave_bad}
\end{figure}

\parTitle{Calculation of non-averaged QMI}
Consider now the QMI $I(\calS:\calE_f)$ for specific environmental fractions $\calE_f$.
To compute the non-averaged QMI, it is necessary to select a specific sequence of increasing environmental fractions, $f\mapsto \calE_f$, to compute the plots of QMI $I(\calS:\calE_f)$ vs $f$.
As we will show, different choices can correspond to radically different behaviours for the QMI. 
Consider three representative scenarios, labeled as A, B, C  in~\cref{fig:ave_bad}.
\begin{itemize}
    \item Scenario A is a ``best-case-scenario'': for all fraction of size $fN \le m$ we only include correlated qubits. For $fN>m$, the remaining qubits are selected arbitrarily from the uncorrelated ones.
    \item In scenario B, we consider instead a ``worst-case-scenario'': every fraction of size $fN \le N-m$ contains exclusively uncorrelated qubits, until $fN>N-m$, at which point all possibly uncorrelated qubits have been selected, and the remaining correlated ones are taken.
    \item In scenario C, every fraction is taken to contain at least one correlated qubit, but at least one of the other correlated qubits is left out, up until $fN=N$, at which point all correlated qubits have necessarily been taken.
\end{itemize}
Interestingly, the mutual information plots for these three different scenarios remain identical, regardless of whether the environment consists of $m=5$ or $m=50$ correlated qubits.
This is because these plots only depend on the ratio between correlated and total qubits, which is fixed in our examples.
The fact that the plots shown in~\cref{fig:ave_bad} are the same both for a case where quantum objectivity emerges ($m=50$), and one where quantum objectivity does not emerge ($m=5$), already shows how non-averaged QMI plots are unable to distinguish between two very different situations. This shows that the non-averaged mutual infomration mostly depends on the ratio between $m$ and $N$, while the average QMI mostly depends on $m$ alone, it was however already showed in~\cite{chisholm2023meaning} how the averaged mutual information quantifies consensus, which has a clear operative interpretation as a quantifier of objectivity.

Scenarios A and B do not display the typical plateau of objective states. However, while the QMI in scenario A is consistently greater than one, in scenario B it is consistently lesser than one.
Therefore in these cases the QMI plots do not allow to reach a definitive conclusion about the presence of objectivity.
In both cases, the QMI exhibits a small atypical plateau, with width exactly $m/N$, which is however very different from the typical plateaus that are expected for objective states, which are symmetric around the $f=1/2$ value.
This suggests that even for the non-averaged QMI, a minimum plateau as large as the redundancy should be present.

Scenario C is more nuanced. A clear objectivity plateau is present, but given that the QMI is $S(\rho_{\mathcal{S}})$ even for a fraction of one qubit, one might overestimate the redundancy as $N$, while only $m$ qubits are actually correlated with the system. Besides grossly overestimating the redundancy, scenario C could also mistakenly suggest that QD-objectivity emerges both when $m=5$ and when $m=50$, which we know to be false from~\cref{fig:ave_good}.

\parTitle{Conclusions}
In more general terms, consensus will be smaller than redundancy whenever there are hindrances to the information recovery.
This happens for example in scenarios where parts of the environment have negligible correlations with the system.
Said hindrances may arise when the information encoding is not local or, as we have seen here, not symmetrical.
In such cases, averaging the QMI becomes paramount, as trying to quantify objectivity via non-averaged QMI can lead to contradicting results, as shown in~\cref{fig:ave_bad}.

For objective states in the macroscopic limit, it is reasonable to assume that several uncorrelated environmental fractions come into play. One might assume that these additional fractions should not modify the degree of objectivity, indeed in our model, QD-objectivity consistently emerges provided that the redundancy is sufficiently large, regardless of the number of uncorrelated qubits.

\section{Random degrees of correlations}
\label{sec:random_correlations}

\begin{figure*}[!]
    \centering
    \subfloat[Flat distribution]{\includegraphics[width=0.45\textwidth]{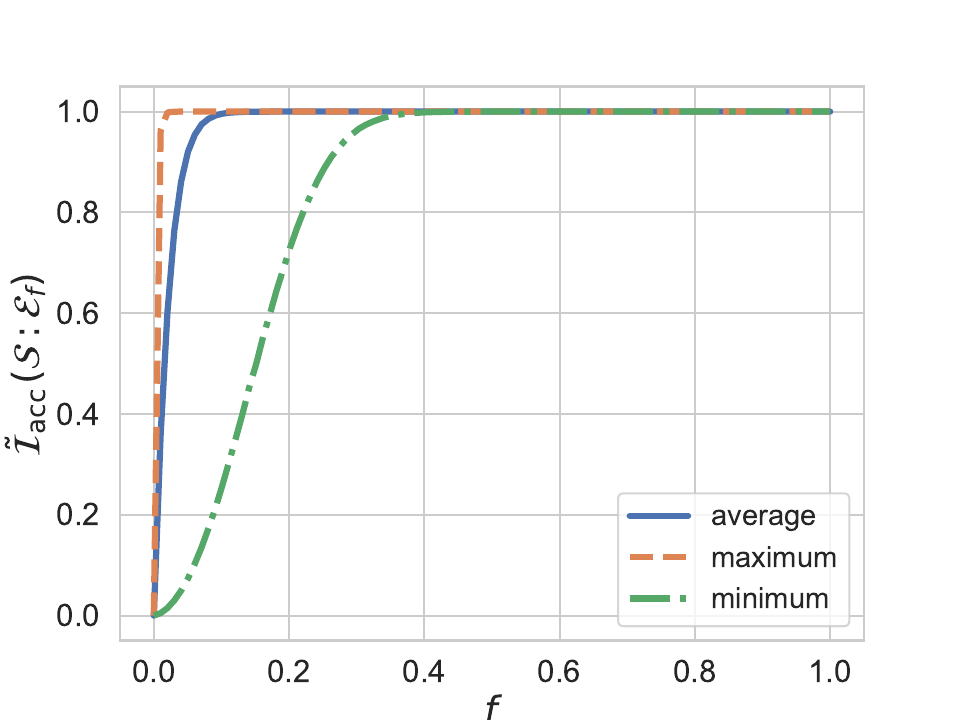}}
    \subfloat[Exponential distribution]{\includegraphics[width=0.45\textwidth]{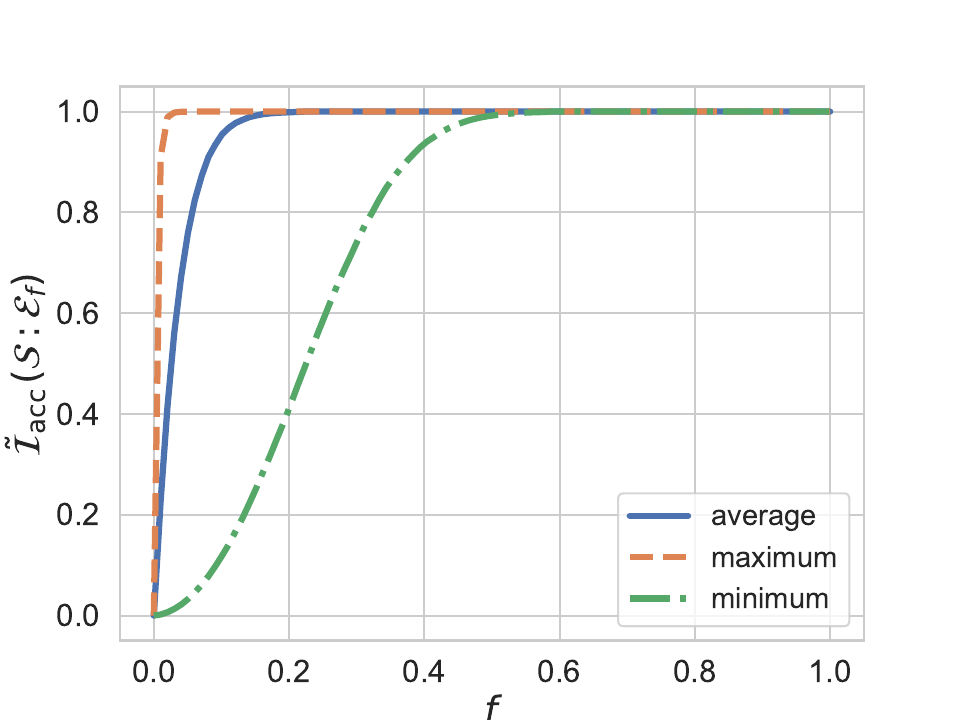}}
    \caption{
    \textbf{Mutual information plots for random degrees of correlation}.
    The environment consists of $N=100$ qubits, and each qubit has a random level of correlations with the system. \textbf{(a)} The correlations for each qubit follow a flat distribution, against a degree of redundancy of $24$, the average mutual information gives a degree of consensus of $11$. We also show the possible results of non-averaged mutual information, which would suggest degrees of consensus of $2$ and $50$ respectively. \textbf{(b)} The correlations follow an exponential distribution, the degree of redundancy is $15$, while the degree of consensus is $6$. Non-averaged mutual information, suggest degrees of consensus of $2$ and $33$ respectively.
    }
    \label{fig:rand_flat}
\end{figure*}

The example in~\cref{sec:asymm_environments} illustrates the importance of using the averaged QMI to witness QD-objectivity in scenarios where a subset of qubits is uncorrelated with the system.
To further stress its importance in a broader context, we consider here scenarios involving random degrees of correlation between system and environmental fractions. 
As we will show, the same conclusions --- namely, the lack of equivalence between redundancy and consensus, and the potential for misleading outcomes when neglecting to average the QMI --- also hold in these more general scenarios.

Let us consider again $(N+1)$-qubit states $\ket\Psi\in\calS\otimes\calE$, where the system $\calS$ interacts with an $N$-qubit environment.
Instead of considering qubits that are either perfectly correlated of totally uncorrelated with the system, consider a case where each environmental qubit interacts with the system via an ``imperfect CNOT'' gate, which we will denote in the following with $i$CNOT.
This is the unitary two-qubit operation that applies the gate $\sqrt{1-p}\sigma_z+\sqrt p\sigma_x$ to the second qubit conditionally to the first one being $\ket1$, with $\sigma_x,\sigma_z$ the standard Pauli matrices.
More explicitly, the $i$CNOT gate has matrix representation
\begin{equation}
i\text{CNOT}=
\begin{pmatrix}
1 & 0 & 0 & 0\\
0 & 1 & 0 & 0\\
0 & 0 & \sqrt{1-p} & \sqrt{p}\\
0 & 0 & \sqrt{p} & -\sqrt{1-p}\\
\end{pmatrix},
\end{equation}
with $p\in[0,1]$.
The terminology $i$CNOT is justified by it collapsing to the standard CNOT for $p=1$.
This type of system-environment interaction is standard when studying objective states resulting from collision models~\cite{lorenzo_anti-zeno-based_2020, touil_eavesdropping_2022}.
The element of novelty here is that for each qubit the value of $p$ is randomly extracted from a probability distribution.

If the initial state is $\ket{+}_\calS\ket{0}^{\otimes N}$, after each environmental qubit has interacted with the system, the state becomes
\begin{equation}%\small
\ket{0}_\calS\ket{0}^{\otimes N}+\ket{1}_\calS\otimes\bigotimes_{i=1}^{N}\left(\sqrt{1-p_i}\ket{0}+\sqrt{p_i}\ket{1}\right),
\label{eq:random_state}
\end{equation}
with each $p_i$ value a random number. 
To simplify our analysis, we will restrict ourselves to measuring system and environmental qubits in the computational basis, which is the basis into which information is encoded, and it is therefore the optimal one for extracting information about the system.
We will therefore compute not the averaged QMI, but the averaged (classical) mutual information. To do so, we will first compute the non-averaged mutual information resulting from the outcome probability of the measurements performed by the observers, which we will refer to as the \emph{accessible} information, $I_{\operatorname{acc}}(\calS:\calE_f)$.

\parTitle{Computation of non-averaged mutual information}
We will first compute the mutual information between $\calS$ and a specific set of $l=fN$ environmental qubits, when all measurements are performed in the computational basis.
We recognize only three possible types of measurement outcomes:
\begin{itemize}
    \item The system is measured in the $\ket{0}_\calS$ state and the environmental qubits are measured in the $\ket{0}^{\otimes l}$ state. This occurs with probability $\frac{1}{2}$.
    \item The system is measured in the $\ket{1}_\calS$ state and the environmental qubits are measured in the $\ket{0}^{\otimes l}$ state. This occurs with probability
	\begin{equation}
	P_f=\frac{1}{2}\prod_i^{fN}(1-p_i).
	\end{equation}
    \item The system is measured in the $\ket{1}_\calS$ state and the environmental qubits are measured in any state that is not $\ket{0}^{\otimes l}$, with probability $\frac{1}{2}(1-P_f)$.
\end{itemize}
Using the probabilities associated with these scenarios, we can compute the marginal probability distribution for the system and for the environmental fraction, and then recover the mutual information
$S_E+S_S-S_{SE}=$
\begin{equation}\label{eq:rand}
\begin{gathered}\small
I_{\operatorname{acc}}(\calS:\calE_f)=\frac{1}{2}\ln(2)+P_f\ln(P_f)
\\
-\left(\frac{1}{2}+P_f\right)\ln(\frac{1}{2}+P_f).    
\end{gathered}
\end{equation}

\Cref{eq:rand} is the \textit{non-averaged} mutual information in the computational basis, and thus depends on the choice of environmental qubits via $P_f$.
On the other hand, the \textit{averaged} mutual information between $\calS$ and $l$ environmental qubits is obtained by extracting $l$ values of $p_i$ from a chosen probability distribution, computing the resulting mutual information, and then averaging over several repetitions of the extraction process. The average is performed over $10^4$ extractions, the results are shown in \cref{fig:rand_flat}.

To highlight even in this scenario the misleading nature of the non-averaged mutual information, we will compute the mutual information with respect to choices of environmental qubits that maximize it or minimize it.
The maximised (minimised) mutual information is obtained by extracting $N$ values of $p_i$ and computing the mutual information with the $m$ highest (lowest) values, which corresponds to the environmental qubits that are most (least) correlated with the environment. We average over $10^4$ extractions, to show that, for typical values of $\{p_i\}$, the resulting biased maximum (minimum) mutual information is higher (lower) than the averaged one.

Consider now the values of redundancy and consensus in this example.
To compute the resulting consensus value we set the threshold value to be $0.99S(\rho_\calS)$, and look for the $f_0$ value such that $\tilde I(\calS:\calE_{f_0})=0.99S(\rho_\calS)$, the consensus value is then given by $1/f_0$.

While the calculation of redundancy performed in~\cref{sec:asymm_environments} for the state in~\cref{eq:first_state} was straightforward, calculating it in more general cases can be nontrivial, requiring to find the largest $n$ such that there is an environmental partition $\calE=\bigotimes_{i=0}^n\calE_i$ 
such that $I(S:E_i)\simeq S(\rho_S)$. This would be computationally demanding in most scenarios.
Here we compute redundancy by simply dividing the environment into fractions, such that the mutual information between each fraction and the system is at least $0.99S(\rho_\calS)$. Also in this case we average over $10^4$ extractions of $\{p_i\}$.
This simple approach is sufficient to provide a lower bound on the redundancy significantly higher than the corresponding consensus, thus matching our initial expectations.

\parTitle{Numerical results}
In~\cref{fig:rand_flat} we show a comparison between averaged, maximized, and minimized accessible information.
Both results are for an environment composed by $N=100$ qubits. 
In~\cref{fig:rand_flat}~(a) the degrees of correlation between the environmental qubits and the system are given by a flat probability distribution, the redundancy is $24$, and consensus is $11$. The non-averaged mutual information plots would instead lead to grossly overestimating or underestimating the degree of objectivity.
Similar results are obtained in~\cref{fig:rand_flat}~(b), where the degrees of correlation between environmental qubits and system are given by an exponential distribution, the redundancy is $15$ while consensus is $6$.
We can see how even in this more realistic scenarios, not performing the average can lead to misleading results.
Furthermore, this case offers yet another example where we can appreciate the numerical difference between consensus and redundancy.

\section{Properties of the non-averaged QMI}
\label{sec:non-averaged-qmi}

The previous section demonstrated explicitly that using the non-averaged QMI can lead to incorrect results, such as false positive or negative witnesses of objectivity, as well as misquantifying redundancy and consensus levels. In this section, we seek to recognize some of the merits of the non-averaged QMI, showing how we can nonetheless extract some intriguing insights from such non-averaged quantities.

\newcommand{\barEf}{\bar\calE_f}

It is a standard result that the averaged QMI $\tilde I(\calS:\calE_f)$ is antisymmetric with respect to $f=1/2$~\cite{blume-kohout_simple_2005}, or more precisely,
\begin{equation}
    \tilde I(\calS:\calE_f) +
    \tilde I(\calS:\calE_{1-f}) =
    2S(\rho_\calS).
\end{equation}
In the case of symmetric information encoding, the averaged QMI reduces to the non-averaged one: $\tilde I(\calS:\calE_f)=I(\calS:\calE_f)$ for any subset $\calE_f$ containing a fraction of $0<f\le1$ environmental qubits.
A similar identity applies for arbitrary encoding scenarios when we do not consider averaged quantities: given any partition $\calE=\calE_f\otimes\barEf$ for some $\calE_f\le\calE$, we have by direct analysis, under the sole assumption that the overall state $\rho_{\calS\calE}$ is pure,
\begin{equation}
    I(\calS:\calE_f)+
    I(\calS:\barEf) = 2S(\rho_\calS).
\end{equation}

From this we see that if for any $f< 1/2$ 
we have
\begin{equation}
    I(\calS:\calE_f)=S(\rho_\calS)+\Delta,
\end{equation}
for some $\Delta>0$, then the information encoding cannot be symmetric, and furthermore
\begin{equation}
    I(\calS:\barEf)=S(\rho_\calS)-\Delta.
\end{equation}
Measuring the whole $\calE_{\overline{f}}$ environmental fraction, i.e. the majority of the environment, still does not allow to recover enough information about the system.

It is interesting to notice how the discordant terms in the QMI provide relevant information even in the context of quantum objectivity.
In particular, witnessing a QMI with an environmental fraction higher than $S(\rho_{\cal S})$ --- which is only possible in the presence of non-zero discord --- puts an upper bound on the mutual information with the rest of the environment.

As highlighted by our discussion in \cref{sec:asymm_environments}, the degree of redundancy can be high even if only a small fraction of the environment is highly correlated with the system.
In other words, it is possible to have small QMI between system and most of the environment, and nonetheless have high redundancy.
On the other hand, if a system is highly correlated with a single environmental qubit, then clearly there is no redundancy.
In conclusion, having a small QMI between system and most environmental fractions is not alone sufficient to infer the degrees of redundancy or consensus.

Witnessing values of $I(\calS:\calE_f)> S(\rho_\calS)$, for small $f$ signals that the system is mostly correlated with only a part of the environment, and that the use of the averaged QMI is needed.
This analysis provides further evidence for the fact that the non-averaged QMI can be a misleading quantifier for quantum objectivity. Notwithstanding, it also shows us that the QMI with respect to a specific environmental fraction provides interesting information about accessible quantities, bounding the possible correlations with respect to the rest of the environment.
This further shows how a quantum system cannot be correlated with QMI higher than the entropy of the system simultaneously with more than one environmental fraction, therefore, objectivity ideally requires that the QMI between the system and the environmental fractions is exactly $S(\rho_\calS)$.
The existence of even a partial objectivity plateau is then a crucial feature of objective states. Since, as previously noted, the width of the objectivity plateau is at least the value of redundancy, the lack of a QMI plateau is sufficient to infer a complete lack of redundancy, and thus of objectivity.

\section{Conclusions}

In this work we highlighted the importance of using the average mutual information when performing measures of quantum objectivity.
In \cref{sec:asymm_environments}, we provided an example with a clear physical interpretation, and showed the possible outcomes of using the non-averaged mutual information, that would result in inaccurate measures of the degree of objectivity. The average mutual information, on the other hand, always offers results with a clear operative meaning. We analysed in \cref{sec:random_correlations} a more generic model, and obtained similar results even in a more complex scenario.
Finally, in \cref{sec:non-averaged-qmi} we discussed some intersting properties of the non-averged QMI, that allow to infer information about the overall structure of the system-environment correlations.
Aside from advocating for the use of the averaged mutual information, this work offers clear interpretational tools for the analysis of (averaged and non-averaged) mutual information plots that can help navigate results for similar scenarios in future works.

\begin{acknowledgments}
% \section{Acknowledgments}
D.A.C.~acknowledges support from the Horizon Europe EIC Pathfinder project QuCoM (Grant Agreement No.101046973). L.I.~acknowledges support from MUR and AWS under project PON Ricerca e Innovazione 2014-2020, “calcolo quantistico in dispositivi quantistici rumorosi nel regime di scala intermedia” (NISQ - Noisy, IntermediateScale Quantum).
\end{acknowledgments}

\bibliography{Ref}

\end{document}